\documentclass[letter]{aa}
\usepackage{graphicx}
\usepackage{txfonts}
\usepackage[Symbol]{upgreek}

\begin{document}

\title{Restarted activity in the 3C\,328 radio galaxy}

\author{A. Marecki}

\institute{Institute of Astronomy, Nicolaus Copernicus University,
           Faculty of Physics, Astronomy and Informatics, Grudzi\k{a}dzka 5,\\
           87-100 Toru\'n, Poland\\
           \email{amr@astro.uni.torun.pl}
}

\date{Received 8 April 2021 / Accepted 20 April 2021}

\abstract{As a rule, both lobes of Fanaroff-Riley (FR) type-II radio sources 
are terminated with hotspots, but the 3C\,328 radio galaxy is a specimen 
of an FR\,II-like object with a hotspot in only one lobe. A conceivable 
reason for such asymmetry is that the nucleus of 3C\,328 was temporarily 
inactive. There was no energy transfer from it to the lobes during the 
period of quiescence, and so they began to fade out. However, under the 
assumption that the axis connecting the two lobes makes an appreciable angle 
with the sky plane, and hence one is considerably farther from the observer 
than the other, the lobes are observed at two distinct stages of evolution 
due to the light-travel lag. While the far-side lobe is still perceived as 
being of the FR\,II type with a hotspot, decay of the near-side lobe is 
already apparent. No jets are visible in the VLA images, but the VLBA 
observations of the inverted-spectrum core component of 3C\,328 have 
revealed that it has a jet of a sub-arcsecond length pointing towards the 
lobe that shows evidence of decay. Since the jet always points to the near 
side, its observed orientation is in line with the scenario proposed here. 
The presence of the jet supports the inference that the nucleus of 3C\,328 
is currently active; however, given the fact that the jet is short 
($\sim$200\,pc in projection), the activity must have restarted very 
recently. The lower and upper limits of the quiescent period length have 
been calculated.
}


\maketitle

\section{Introduction}

The intermittent nature of active galactic nuclei (AGNs) can manifest itself 
in a number of ways. A direct observable indication of the cessation and 
subsequent restart of activity in a~radio-loud AGN takes the form of a 
double-double radio source \citep[DDRS;][]{Schoenmakers2000} consisting of 
two co-linear pairs of radio lobes (see \citealt{Saikia2009} for a review 
and e.g. \citealt{Marecki2021} for references to the latest papers on 
DDRSs). If the inner lobes resulting from the second active phase of a DDRS 
had been created only recently, their separation could have been too small 
to render them as a double in the image depicting the overall structure of 
the source. It follows that as long as the putative `core' is not resolved, 
the true nature of this kind of object remains unknown. B\,0818+214 is a 
good example demonstrating such a case. Only MERLIN and the EVN observations 
revealed that its central component, which is straddled by a pair of 
large-scale lobes as seen in the Faint Images of the Radio Sky at 
Twenty centimeters (FIRST) survey \citep{Becker1995}, is actually the inner 
double that is two orders of magnitude more compact than the co-linear outer 
double \citep{Marecki2009}. Therefore, although seemingly a~core-dominant 
triple (CDT), B\,0818+214 is also a DDRS.

However, a conjecture that every CDT is a concealed DDRS, such as is the 
case of B\,0818+214, would be misguided. \citet[][hereafter 
Paper\,I]{Marecki2011} studied three quasi-stellar objects (QSOs) that are 
CDTs whose lobes are asymmetric: While a hotspot is present in one of the 
two lobes of each of these QSOs, the other is diffuse and lacks a hotspot. 
According to the authors, the cause of this phenomenon is as follows. The 
activity of the nuclei in those objects has ceased, and hence the energy 
supply to the lobes has been cut off. As a consequence, as the lobes weaken 
and become diffuse, the hotspots vanish. If the axis of a double-lobed radio 
source lies far from the sky plane, which is the case for QSOs 
\citep{Barthel1989}, then the perceived epoch of the far-side lobe is 
considerably earlier than the perceived epoch of the near-side lobe due to 
the light-travel time \citep{Gopal1996}. For large-scale sources, the 
difference between those epochs may be of the order of at least 
$10^5$\,years. Since the lifetimes of the hotspots are $7\times10^4$\,yr 
\citep{Kaiser2000}, such light-travel time is sufficient for the hotspot of 
the near-side lobe to disappear while the hotspot of the far-side lobe is 
still observable. This kind of the lobe asymmetry can thus be regarded as 
a~signature of a recent termination of activity. Indeed, the EVN 
observations reported in Paper\,I show that the cores of those CDTs are 
typical for radio-quiet AGNs \citep{Ulvestad2005}.

\begin{figure*}
\centering
\includegraphics[width=0.47\linewidth]{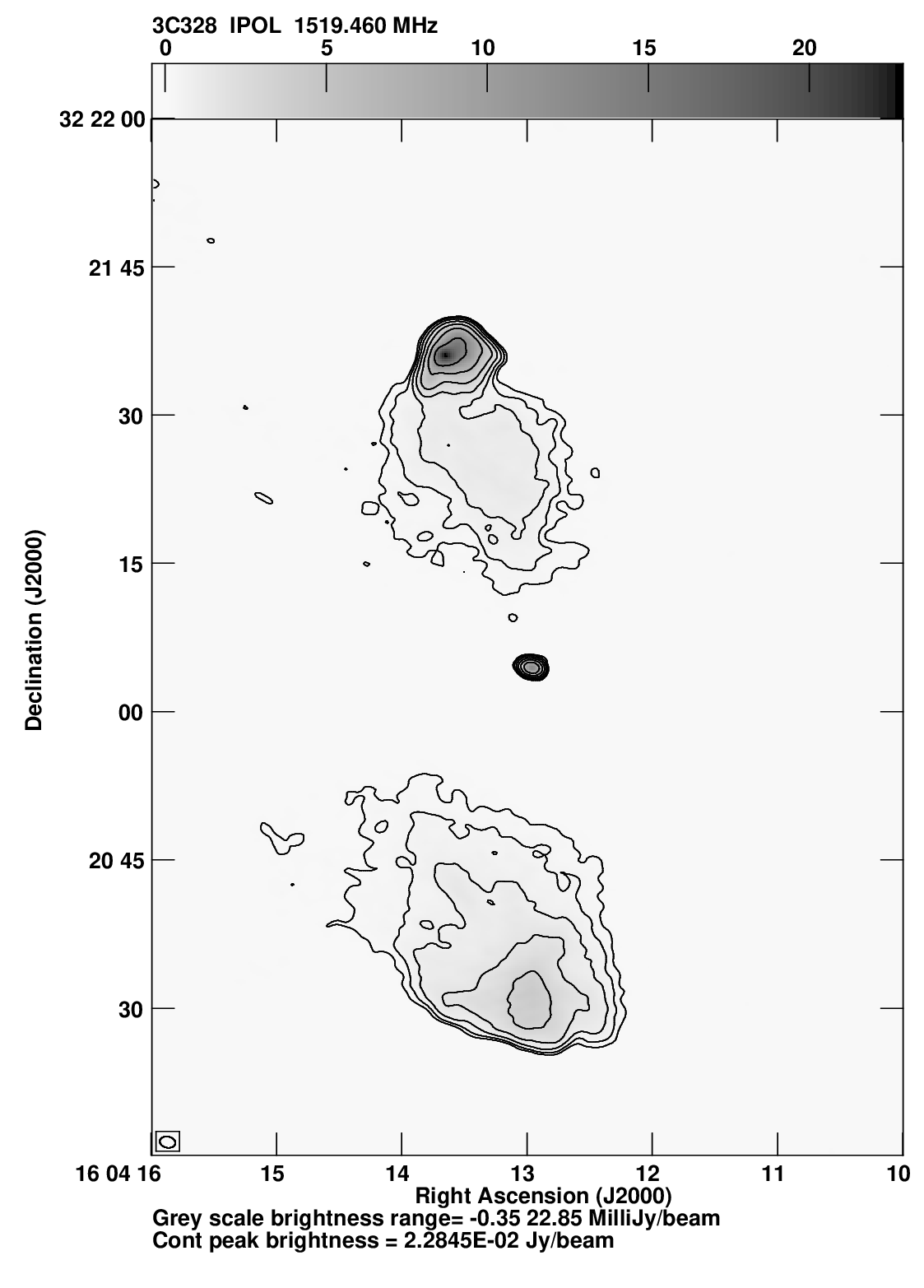}
\includegraphics[width=0.47\linewidth]{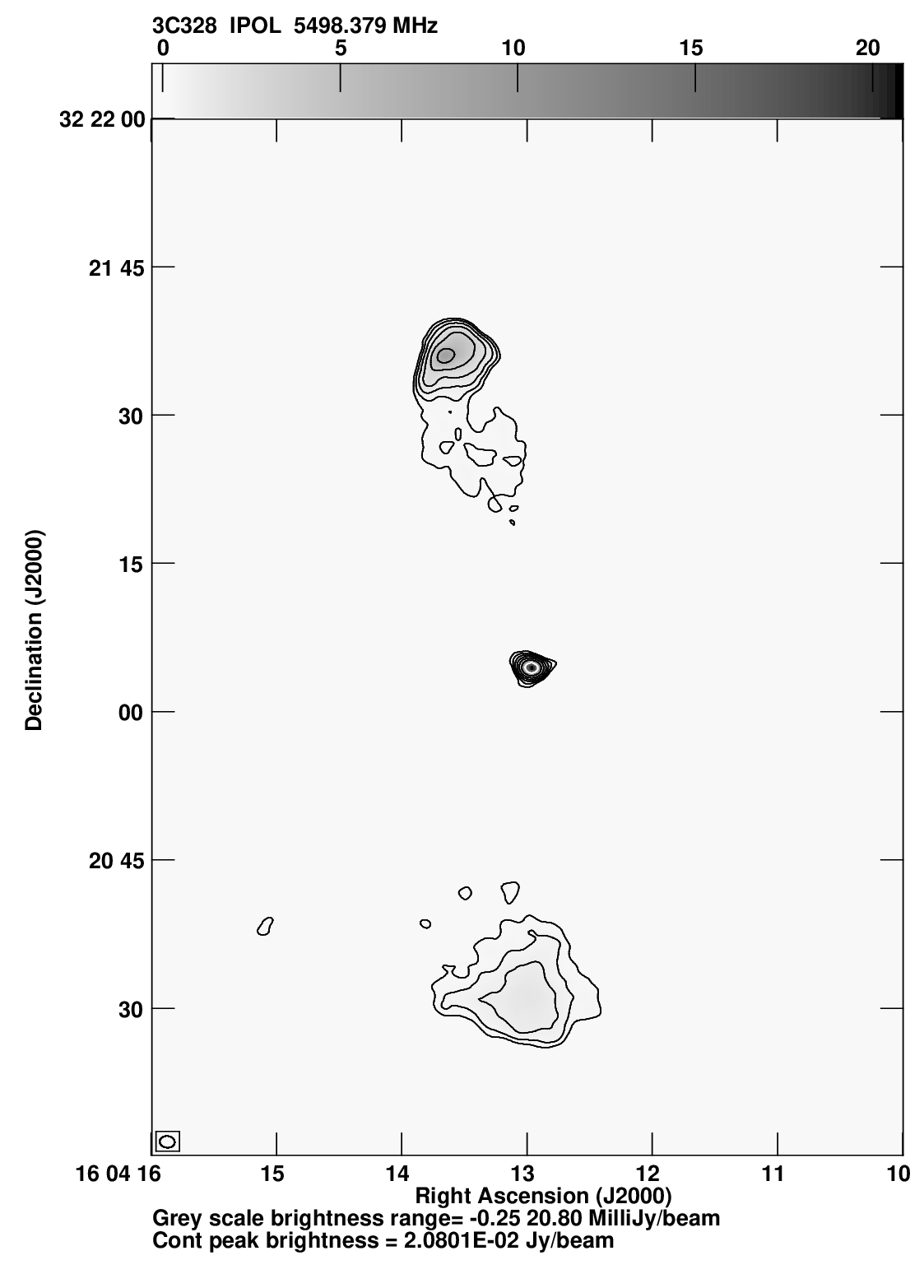}
\includegraphics[width=0.43\linewidth]{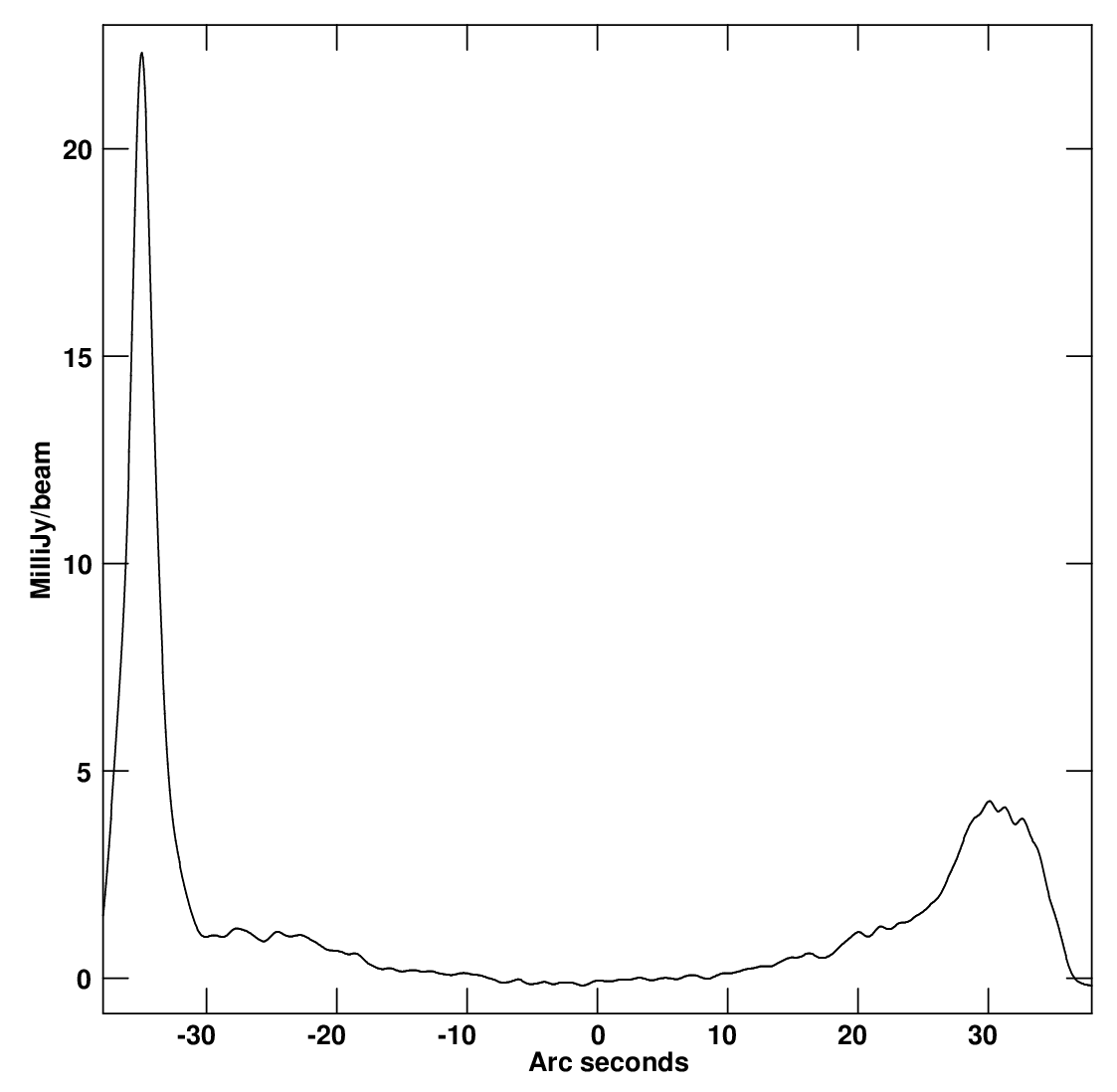}
\hspace{0.7cm}
\includegraphics[width=0.43\linewidth]{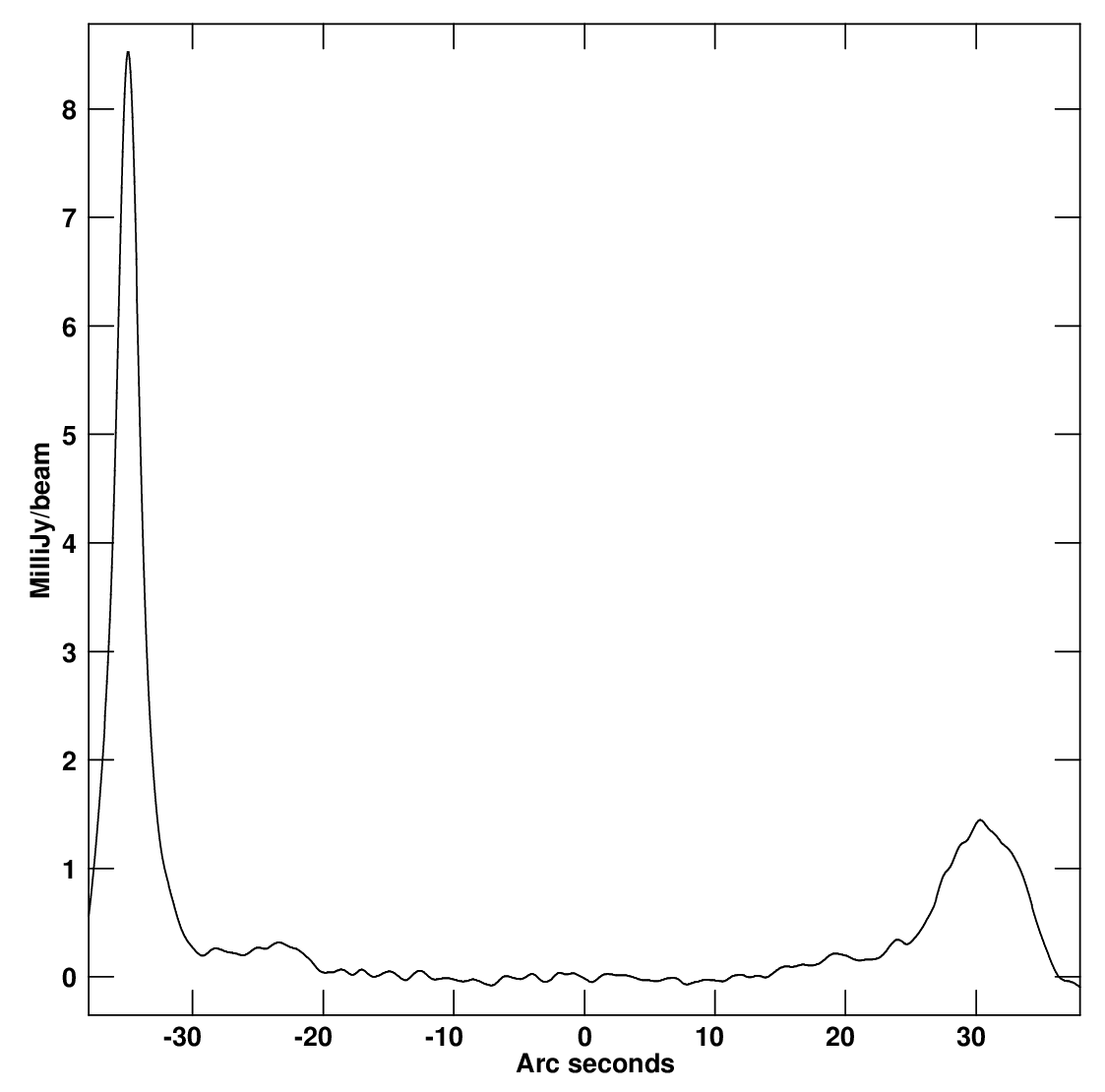}
\caption{Outcome of the VLA observations.
{\it Upper panels}: Images of 3C\,328 at 1.5\,GHz ({\it left-hand panel}) 
and at 5.5\,GHz ({\it right-hand panel}). In both images, contours are at 0.2, 
0.4, 0.8, 1.6, 3.2, 6.4, 12.8, and 25.6\,mJy/beam. {\it Lower panels}: Slices 
across the lobes at 1.5\,GHz ({\it left-hand panel}) and at 5.5\,GHz ({\it right-hand 
panel}). North is on the left-hand side of each plot. The core has been 
bypassed by both slices.}
\label{fig:VLA}
\end{figure*}

\begin{figure*}
\centering
\includegraphics[width=0.49\linewidth]{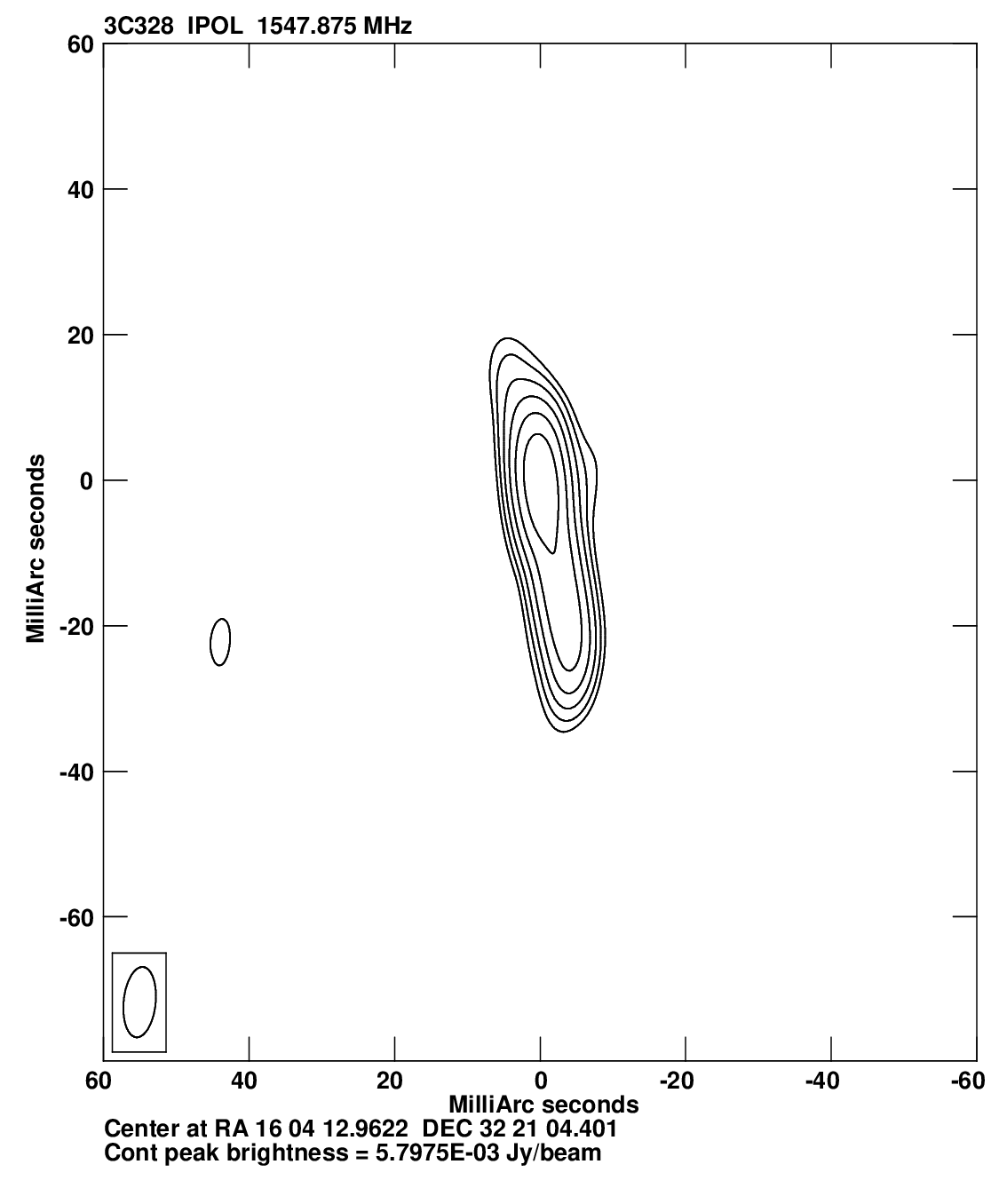}
\includegraphics[width=0.49\linewidth]{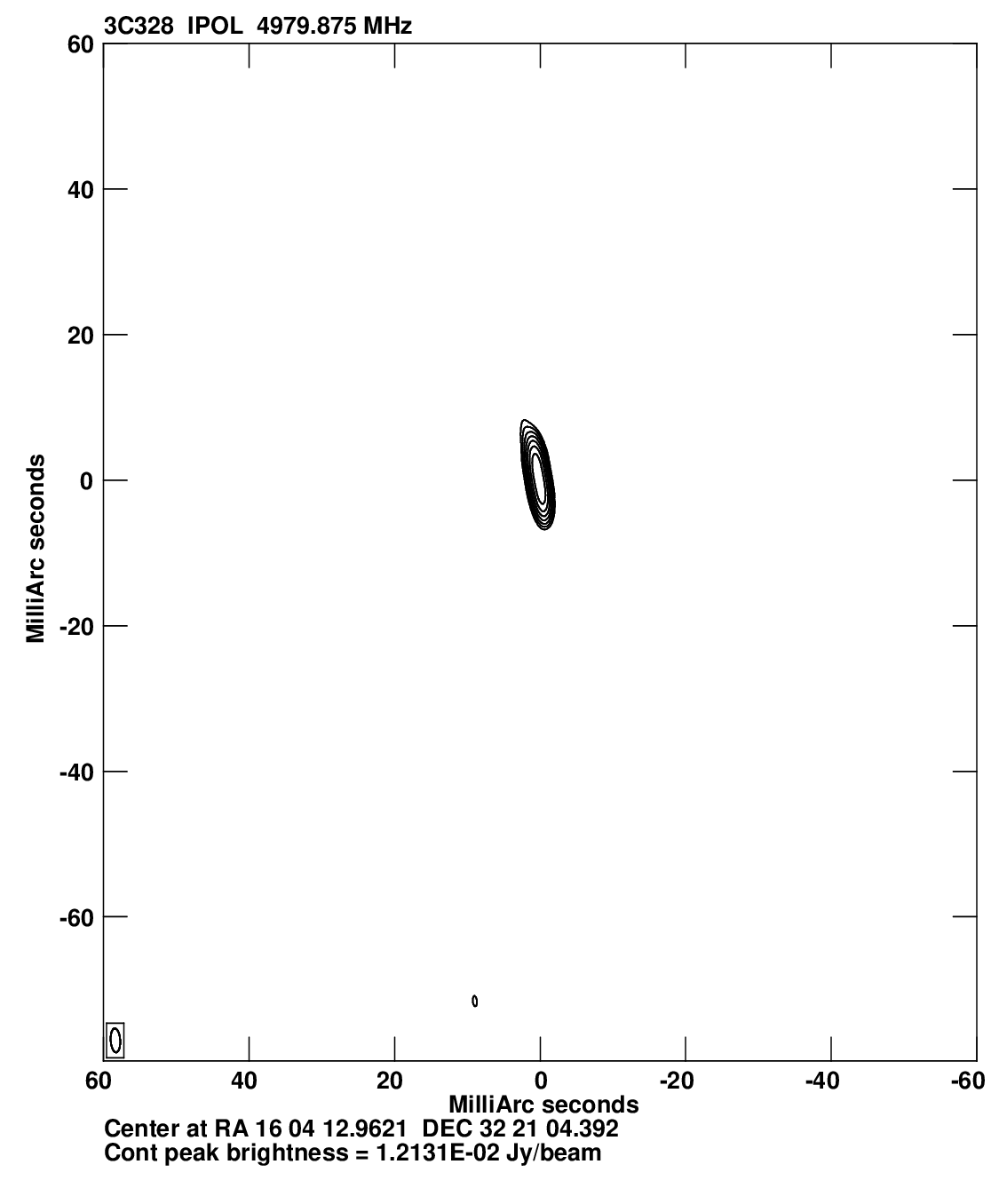}
\caption{VLBA images of 3C\,328 at 1.5\,GHz ({\it left-hand panel}) and at 5\,GHz 
({\it right-hand panel}). In both images, contours are at 0.1, 0.2, 0.4, 0.8, 1.6, 
3.2, 6.4, and 12.8\,mJy/beam.}
\label{fig:VLBA}
\end{figure*}

By no means is the lopsidedness of the lobes highlighted above limited to 
QSOs. It is just as likely to take place in a radio galaxy, unless such a 
galaxy lies close to the sky plane, provided that its lobe-to-lobe span is 
sufficiently large for the aforementioned time lag to become significant. 
For example, the morphology of the \object{3C\,16} radio galaxy can be 
explained within the framework of that scenario. It is asymmetric in the 
same manner: Whereas the south-western lobe is of a Fanaroff-Riley (FR) 
type\,II \citep{Fanaroff1974}, the north-eastern one is devoid of a hotspot 
(see the images in \citealt{Leahy1991}, \citealt{Harvanek1998}, and 
\citealt{Gilbert2004}). The nucleus of 3C\,16 is not just radio quiet; given 
that there is no radio core visible, it is fully radio silent. As the 
activity has ceased in this object, it could be labelled `post-active'.

Post-active objects can, however, revert to the active state. Active 
galactic nuclei such as \object{J1211+743} (see the image in 
\citealt{Pirya2011}) as well as \object{3C\,249.1} and \object{3C\,334} (see 
the images in \citealt{Gilbert2004}) have restarted. These three AGNs are 
unique for two reasons. Firstly, they evince the asymmetry of the type 
described above, and secondly their jets are oriented towards the lobe with 
no hotspot. As posited by the light-travel-time-based scenario, the lobe 
showing evidence of decay must be on the near side. Since the jet always 
signposts the near side, such an orientation of the jet supports that model. 
Detailed studies of these three sources are presented in \citet[][hereafter 
Paper\,II]{Marecki2012a}, where the model was introduced and applied to 
J1211+743, as well as in \citet[][hereafter Paper\,III]{Marecki2012b}, where 
it was applied to 3C\,249.1 and 3C\,334.

Radio galaxy \object{3C\,328} (B2\,1602+324) is also a source that has one 
FR\,II-type lobe and whose other lobe has no hotspot (see the VLA image at 
1465\,MHz in \citealt{Machalski1983}). Moreover, there is a~point-like core 
at the centre, so the morphology of 3C\,328 very much resembles that of the 
QSOs from Paper\,I despite the fact that it is not a QSO but rather a 
galaxy. Whether it has a switched-off nucleus like those QSOs or is 
restarted is not known. To investigate the current state of activity of the 
nucleus of 3C\,328, the VLA and VLBA observations have been carried out, and 
their outcome is reported here. The following values of cosmological 
parameters are used in this Letter: $H_0\!=\!67.4\,{\rm km\,s}^{-1}{\rm 
Mpc}^{-1}, \Upomega_{\rm M}\!=\!0.315$, and $\Upomega_{\Uplambda}\!=\!0.685$ 
\citep{Cosmo2020}; hence, for the redshift of 3C\,328, $z\!=\!0.45243$, 
1\,arcsec of the angular size translates to 5.971\,kpc.

\section{Observations}
\label{sect:obs}

The VLA observations were performed in two configurations: `B' at two 
1024-megahertz-wide frequency bands centred at 5\,and 6\,GHz on 
11~March~2019, and `A' at two 512-megahertz-wide frequency bands centred at 
1264 and 1776\,MHz on 9 October~2019. In each of those two runs, 3C\,328 was 
observed for 10\,min. J1606+3124 was used as a phase calibrator. Continu\-um 
data reduction was carried out using the VLA \textsc{casa} calibration 
pipeline. The 1.5\,GHz images were generated with Briggs weighting, which 
yields a good trade-off between resolution and sensitivity. The same 
weighting was initially used for the 5.5\,GHz data, but the diffuse features 
of the object were poorly rendered in the image. Therefore, the natural 
weighting that yields the highest signal-to-noise ratio at the cost of a 
poorer angular resolution was eventually applied. After adopting this 
scheme, the restoring beam sizes at both frequencies were very similar: 
$1\farcs59\times1\farcs12$ at 1.5\,GHz and $1\farcs52\times1\farcs18$ at 
5.5\,GHz.

The VLBA observations were performed on 8~June~2013 at two 256-megahertz-wide 
frequency bands centred at 1552\,MHz and 4980\,MHz. In each of these two 
bands, 3C\,328 was observed for 100\,min. The restoring beam sizes were 
$9.70\!\times\!4.40$\,mas at 1.5\,GHz and $3.29\!\times\!1.30$\,mas at 5\,GHz.

\begin{table}[ht]
\centering
\caption{Flux densities of the components of 3C\,328.}
\begin{tabular}{c c c c c}
\hline
\hline
Frequency & \multicolumn{4}{c}{Flux densities [mJy]}\\
	    \cline{2-5}
{[MHz]}   & Northern & Southern & Northern & Core \\
          & lobe     & lobe     & hotspot  &      \\
\hline
1519 & 264 & 242 & 107 & 12.2 \\
5498 &  71 &  51 &  34 & 21.4 \\
\hline
\end{tabular}
\label{tab:fluxes}
\end{table}

\section{Results}

The VLA images resulting from the observations mentioned in 
Sect.\,\ref{sect:obs} are presented in the upper panels of 
Fig.\,\ref{fig:VLA}. While the hotspot is present in the northern lobe, the 
absence of a hotspot in the southern lobe is conspicuous, particularly at 
5.5\,GHz (upper right-hand panel of Fig.\,\ref{fig:VLA}). The dominance of 
the only hotspot is highlighted by the slices running across both lobes but 
bypassing the core (lower panels of Fig.\,\ref{fig:VLA}). The flux densities 
of the lobes measured with \texttt{TVSTAT} \textsc{aips} Utility as well as 
those of the core component and the hotspot of the northern lobe -- both 
measured with \texttt{JMFIT} \textsc{aips} Task -- are displayed in 
Table\,\ref{tab:fluxes}. A compelling trait of 3C\,328 revealed by the data 
shown there is that the core is stronger at 5.5\,GHz than at 1.5\,GHz, which 
means that it has an inverted spectrum ($\alpha_{1.5}^{5.5}\!=\!0.44$, 
$S\!\propto\!\nu^\alpha$) between these two frequencies. This feature has 
been confirmed by the measurement of the flux density at the intermediate 
frequency of 3\,GHz. To this end, the image of 3C\,328 was extracted from 
the \textit{VLA Sky Survey} (VLASS)\footnote{The science case, the 
observational strategy for VLASS, and the results from early survey 
observations are presented by \citet{Lacy2020}.} for Epoch 2.1 with the 
CIRADA cutout service\footnote{http://cutouts.cirada.ca/}, and the flux 
density of the core was measured with \texttt{JMFIT}. It amounts to 
15.9\,mJy, and the following values of the spectral indices were obtained: 
$\alpha_{1.5}^{3.0}\!=\!0.39$ and $\alpha_{3.0}^{5.5}\!=\!0.49$. The 
observations that led to the flux-density measurements at those three 
frequencies were carried out in similar epochs, and therefore the impact of 
the potential long-term variability of the core of 3C\,328 should not be 
significant. It seems that such a variability does indeed take place there. 
According to the FIRST survey, the 1.4\,GHz flux density of the core of 
3C\,328 as observed about two decades earlier amounts to 17.4\,mJy. It 
appears then that the flux density of the core, at least at frequencies 
around 1.5\,GHz (see Table\,\ref{tab:fluxes}), has been on the decline ever 
since.

Based on the outcome of the two-frequency VLA observations, the 
inverted-spectrum core must be self-absorbed, and hence very compact. Its 
VLBA images resulting from the observations mentioned in 
Sect.\,\ref{sect:obs} are shown in Fig.\,\ref{fig:VLBA}. At 5\,GHz, the core 
of 3C\,328 is point-like, while at 1.5\,GHz it has an additional feature 
elongated southwards. As the same phase reference was used throughout those 
observations (J1606+3124), the absolute positions of the source at both 
frequencies were preserved. The core visible at 5\,GHz is co-incident with 
the brightest part of the source in the 1.5\,GHz image, and thus this image 
depicts a core-jet structure where the jet points to the southern lobe. This 
means that, apart from the lobe asymmetry, there is another similarity that 
3C\,328 shares with J1211+743, 3C\,249.1, and 3C\,334: All these AGNs have 
jets that point towards the lobe that is devoid of a hotspot. This raises 
the question of whether the light-travel-time-based model elaborated on in 
Paper\,II for J1211+743 and then used to interpret 3C\,249.1 and 3C\,334 in 
Paper\,III could also be applied to 3C\,328. To find that out, one has to 
perform a quantitative test using that model since it imposes tight 
constraints on both the lower and upper limits of the length of the period 
of quiescence, $t_{\rm q}$.

The projected angular far-side and near-side arm lengths of 3C\,328 
(35\farcs6 and 38\farcs8) respectively translate to 212\,kpc and 232\,kpc of 
the projected linear size, whereas the projected angular size of the jet 
(0\farcs035) translates to 209\,pc of the projected linear size. For these 
linear sizes, the lower ($t_{\rm q}^{\rm min}$) and upper ($t_{\rm q}^{\rm max}$) limits of 
$t_{\rm q}$ were calculated using Eqs.\,(2) and (4) of Paper\,II for 
$\beta_{\rm adv}=0.35$, three values of $\beta_{\rm jet}$ (0.8, 0.9,~and\,1), and 
five values of the angle between the line of sight and the AGN axis 
$\theta$. Since 3C\,328 is a galaxy and not a QSO, only the values 
$\theta\ge45\degr$ are applicable to it \citep{Barthel1989}. The results are 
shown in Table\,\ref{table:limits}. Due to the small size of the jet in 
3C\,328, the calculated limits are almost independent from the adopted value 
of $\beta_{\rm adv}$. The definitions of $\beta_{\rm jet}$ and $\beta_{\rm adv}$ are 
provided in Paper\,II.

\begin{table}
\centering
\caption{Lower and upper limits to the length of the quiescent period.}
\begin{tabular}{c r r r r r r}
\hline
\hline
$\theta$ & \multicolumn{3}{c}{$t_{\rm q}^{\rm min}$ [Myr]}
         & \multicolumn{3}{c}{$t_{\rm q}^{\rm max}$ [Myr]}\\
           \cline{2-4} \cline{5-7}
$[\degr]$ & $\beta_{\rm jet} = $0.8 & 0.9 & 1.0 & $\beta_{\rm jet} = $0.8 & 0.9 & 1.0\\
\hline
45 & 0.65 & 0.50 & 0.38 & 1.91 & 1.78 & 1.67 \\
50 & 0.67 & 0.53 & 0.42 & 1.71 & 1.58 & 1.48 \\
60 & 0.72 & 0.60 & 0.50 & 1.39 & 1.28 & 1.20 \\
70 & 0.80 & 0.69 & 0.60 & 1.17 & 1.07 & 0.99 \\
80 & 0.89 & 0.79 & 0.70 & 1.00 & 0.90 & 0.82 \\
\hline
\end{tabular}
\label{table:limits}
\end{table}

\section{Conclusion}

Taking into account that for at least two restarted AGNs -- \object{3C\,293} 
\citep{Machalski2016} and \object{J1835+6204} \citep{Konar2012} -- the values
of $t_{\rm q}$ also stay within the limits shown in Table\,\ref{table:limits}, it
is plausible that the above limits for $t_{\rm q}$ may be valid for 3C\,328, and as 
such the light-travel-time-based scenario can be used to explain its nature. 
It follows that the activity of 3C\,328 has restarted and, given the minute 
size of its jet, that this happened recently. The images presented here are 
thus a record of two subsequent transitions in the nucleus of 3C\,328: 1) 
from the active to the inactive state, which is expressed by the asymmetry 
of the lobes, and 2) back to the active state, represented by the 
sub-kiloparsec-scale jet. The bottom line is that, in terms of the stage of 
the recurrent activity, the eight asymmetric objects mentioned in this Letter 
can be arranged in chronological order:

\begin{enumerate}
\item The objects studied in Papers\,II and\,III restarted a relatively
long time ago, which has allowed their jets to attain large sizes.
\item Galaxy 3C\,328 has recently restarted after a period of quiescence.
\item Galaxy 3C\,16 is quiescent.
\item The objects described in Paper\,I have just switched off.
\end{enumerate}

\begin{acknowledgements}
The National Radio Astronomy Observatory is a facility of the National 
Science Foundation operated under cooperative agreement by Associated 
Universities, Inc.
This research has made use of the CIRADA cutout service at URL 
cutouts.cirada.ca, operated by the Canadian Initiative for Radio Astronomy 
Data Analysis (CIRADA). CIRADA is funded by a grant from the Canada 
Foundation for Innovation 2017 Innovation Fund (Project 35999), as well as 
by the Provinces of Ontario, British Columbia, Alberta, Manitoba and Quebec, 
in col\-laboration with the National Research Council of Canada, the US 
National Radio Astronomy Observatory and Australia's Commonwealth Scientific
and Industrial Research Organisation.
This research has made use of the NASA/IPAC Extragalactic Database (NED), 
which is funded by the National Aeronautics and Space Administration and 
operated by the California Institute of Technology.
I am grateful to Marek Jamrozy for critical reading of the draft of this
Letter and a number of suggestions that improved it.
\end{acknowledgements}

{}

\end{document}